# Dynamic response of ferrofluidic deformable mirrors using elastomer membrane and overdrive techniques

Maxime Rochette[a], Ermanno F. Borra[a*], Jean-Philippe Déry[b] and Anna M. Ritcey[b]

[a]*Département de physique, de génie physique et d'optique and Centre d'optique, photonique et laser (COPL), Université Laval, Quebec city, Canada;*

[b]*Département de chimie, Université Laval, Quebec city, Canada*

*Ermanno F. Borra, e-mail: borra@phy.ulaval.ca

# Dynamic response of ferrofluidic deformable mirrors using elastomer membrane and overdrive techniques

Here are presented the experimental results obtained with a ferrofluidic deformable mirror controlled by electro-magnet actuators. Using a step input through a single actuator we obtain a steady-state settling time of 100ms; however, different combinations of overdrive inputs can be used to decrease it to 25ms. Also, a new technique which consists in laying down an elastomer membrane, coated with an aluminum film, on the ferrofluid is discussed. By adding the membrane on the ferrofluid, it further decreases the time response by a factor of 2. Furthermore, the thin aluminum layer improves the reflectivity of the mirror. Finally, using the membrane and the overdrive techniques combined, the time response is improved by a factor of 20. Numerical simulations show that ferrofluidic mirrors using membranes and improved electronics should reach settling times of the order of a millisecond. Presumably even lower settling times could be possible.

**Introduction**

During the past few years, our team has carried out research in adaptive optics using deformable ferrofluidic mirrors [1-5]. Although the demonstration of a novel type of deformable mirror that can generate high amplitude deformations with RMS residuals of $\lambda/10$ was made, there were issues that needed to be addressed. According to Parent et al. [2], the ferrofluid used for the mirror (EFH1, a low cost and not too viscous ferrofluid), has a cutoff frequency of about 10 Hz, which is below the requirements for many applications. Nevertheless, this promising work showed that much higher cutoff frequencies could be achieved by changing the shapes of the driving pulses (see Figure 1) but it still needed to properly quantify the wavefront amplitudes.

With the acquisition of new equipment, we had to develop new controllers to obtain similar results and to quantify the amplitudes. After carrying out numerical simulations we obtained an improved overdrive controller. Reflectivity is also an important issue, since the ferrofluid only has 4% reflectivity, which would prevent any astronomical or ophthalmic applications [9]. To tackle this problem, computer simulations were carried out and laboratory experiments were made using a ferrofluid covered with an elastomer membrane and an aluminum coating. The final goal was to have an adaptive optical mirror with a high refresh rate, up to a few kHz and a sufficient reflectivity (Al as a 90% reflectance in the visible spectrum).

**Method**

Application of a current to the electro-magnet actuators, used to shape the surface of the deformable mirror, generates a magnetic field that the ferrofluid reacts to. However, the deformation is not instantaneous and it is critical to know the way in which the surface responds over time when applying corrections to compensate for wavefront aberrations. The time response of the transfer function which makes the link between the input and the output of a system is given by Eq. (1)

$$y(t) = x(t) * g(t) \qquad (1)$$

where y(t) is the output, x(t) the input and g(t) the transfer function.
Fundamental to control theory is the requirement that the settling time (typically the time when the signal is within ±5% of the target) must not depend on the target amplitude [8]. This means that if the system has to reach a higher amplitude, it will need a higher initial velocity. However, in our case, it does indirectly, because of the physical limits of our present equipment. The hardware cannot generate an infinite current, thus limiting our overdrive signal. Nevertheless, this can be used to our advantage as long as we don't refer only to the settling time as the only important parameter to analyze stability and target error. It is then possible to use this greater speed to reach lower amplitudes faster by overdriving the actuators. The overdrive will be a shorter pulse, positive or negative that will increase the time response of the liquid (see Figure 1).

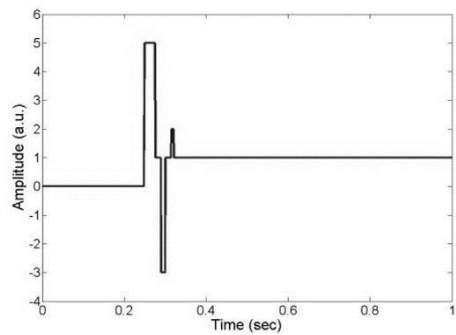

Figure 1. Typical series of pulses used to stabilize the liquid as fast as possible.

For the purpose of this work, a new optical bench using a Shack-Hartmann sensor at 395Hz was acquired which can also be used in a woofer-tweeter configuration [6]. The mirror (Figure 2), custom made by ALPAO, has 91 actuators independently fed with a PD2-AO-96-16 96/16-bit analog output channels PCI card from United Electronic Industries. The actuators are located in a hexagonal pattern so that they fill most of the space within the pupil. For our purpose, the mirror was most of the time used with a 37 actuators configuration within a pupil of 30mm diameter, because it was easier to use since it had less degrees of freedom and needed less computing power. Using only 37 actuators was to demonstrate the principle and no decrease in performances is expected using 91 actuators. Prior results [1] were obtain using the full 91 actuators with very good quality. The core-to-core distance between the actuators is 5mm and their resistance is 2.8Ω. The actuators had 200 turns mounted on a brass frame. The actual limitation resides in a current of 0.2A that can be inputted in the actuators as going higher than this would create enough heat for convection current to establish in the ferrofluid, adding some turbulence to the system. The PCI card doesn't allow a refreshing rate of more than 3kHz, limiting the speed performances.

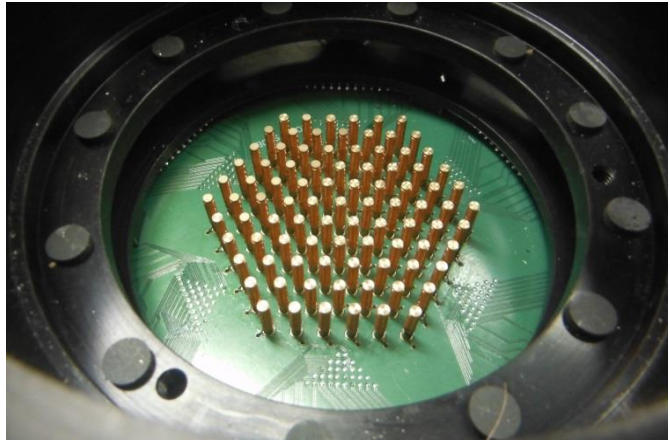

Figure 2. The 91 actuators deformable mirror currently used. It is mainly used in a 37-actuator configuration within a pupil of 30mm.

To optimize the settling time, one can also add an elastomer membrane which, furthermore, allows the mirror to have a good reflectivity by depositing a thin layer of aluminum on its surface, as shown on Figure 3.

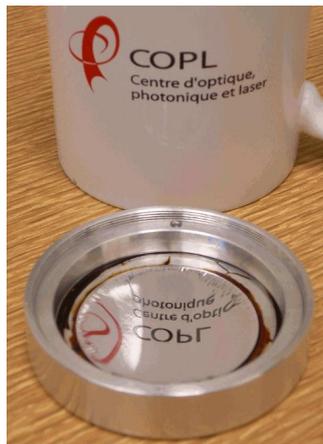

Figure 3. Ferrofluidic mirror covered with an elastomer membrane coated with a thin layer of aluminum.

**Experimental and computer simulation results**

The following section presents results in order of complexity towards decreasing the settling time; starting with results for a single actuator, followed by results on the deformation of a complete surface and finally the development toward using a reflective membrane is presented.

The surface characterization of ferrofluidic mirrors has been reported in [1] and will not be repeated here, the focus of the present work being the time response of the mirror. Figure 4 shows the time response to a 1.5µm amplitude step input for a single actuator.

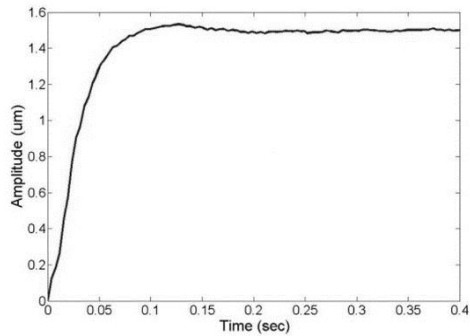

Figure 4. Experimental step response of a single actuator using EFH1 ferrofluid. The settling time is about 100ms. This is a single measurement, not averaged.

Stabilization occurs at about $100\ ms$ with the settling time defined as the time when the system reaches and stays within $\pm 5\%$ of the desired target. Therefore, without a controller, the system cannot be refreshed more often than 10 times per second since it needs to reach stability before applying another correction. Using an overdrive controller that generates the series of pulses shown in Fig. 1 decreases significantly this settling time, as shown in Figure 5 and Figure 6.

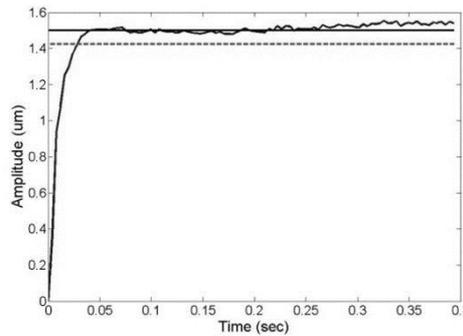

Figure 5. Deformation amplitude of an overdrive command for a single actuator using EFH1 ferrofluid. The dotted line gives the $-5\%$ error margin. This is a single measurement, not averaged.

Figure 5 shows that with the right combination of overdrive pulses (like those shown in Figure 1) determined with simulations and/or experimentally it is possible to reduce the settling time by a factor of 4 with respect to the 100 ms of a single pulse (Figure 4). In Figure 5, the maximum overdrive lasts 6.5 ms and has 10 times the amplitude of the actual target. It is then followed by a negative drive lasting 1.5ms and having 6 times the amplitude of the actual target. Optimization of the pulse sequence is not trivial since, although one can come close with computer simulations, some fine tuning is also required in the laboratory.

As explained earlier, the settling time does not depend on the target amplitude. To achieve shorter settling times, one must increase the difference between the overdrive command and the target command. Because we are limited, with our present equipment, in the current that the actuators can sustain (not to generate too much heat and thereby create convection currents in the ferrofluid), we can, in practice, reach low target amplitudes faster than higher ones. It is, however, possible to simulate a very high ratio of overdrive and target commands which could reduce the settling time to as little as 5 ms (see Figure 6). Unfortunately, this speed cannot be experimentally implemented with current equipment.

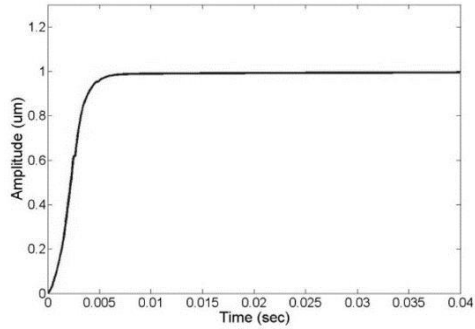

Figure 6. Computer-simulated response of a single actuator using the transfer function of the system with a very high ratio (100 times) of overdrive and target command. The settling time is $5 ms$.

The transfer function of the system used in Figure 6 is derived from experimental results shown in Figure 4. By using Eq. 1, we fitted this output and divided its Laplace transform by the Laplace transform of the inputted step function. The inverse Laplace transform of the latter division lead to the transfer function of the system which could be used to generate computer-simulated responses.

Despite the experimental speed limitation, it is relevant to determine how much amplitude can be reached at the optimal speed of 25ms. If the amplitude in Figure 5 is doubled, the ratio of command is reduced by the same factor. So, with the actual setup, it is interesting to know what is the highest reachable amplitude while keeping the $25\ ms$ settling time shown in Figure 5.

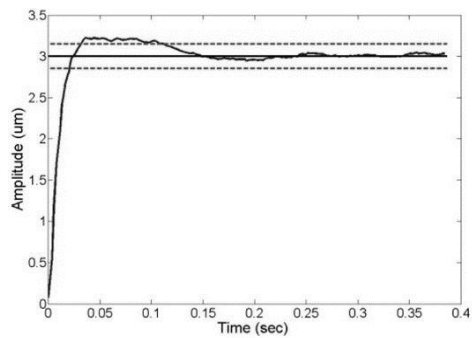

Figure 7. Experimental result of an overdrive command for a single actuator using EFH1 ferrofluid with the target amplitude doubled. The dotted lines are the $\pm 5\%$ margin. This is a single measurement, not averaged.

Figure 7 shows that it is possible to obtain the same settling time of $25\ ms$ with a target amplitude twice as large as the one in Figure 5. Of course, the parameters of the overdrive controller changed since it was not the same target and the maximum ratio between the overdrive and target command was only 5. It is quite encouraging but higher than $3 \mu m$, it was impossible to obtain the same settling time with our present equipment.

After having done the characterization for a single actuator, the next step was to do it for the whole surface, using Zernike polynomials [7] as references to create the target. However, we find that the settling time response is not the same for all Zernike polynomials. Consequently, the settling time without the use of a controller is different from one surface to another because of the different spatial frequencies. Figure 8 shows the time response of a defocus using 37 actuators.

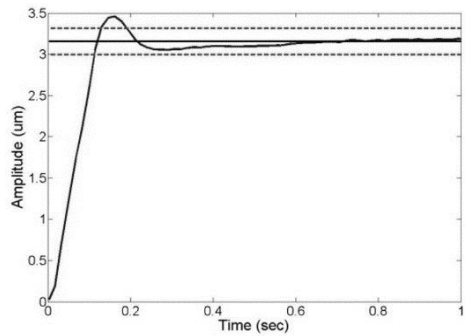

Figure 8. Experimental result of a step response creating the aberration of defocus in the 37-actuators configuration. The dotted lines give the $\pm 5\%$ margin. The settling time is around $200 ms$. This is a single measurement, not averaged.

Using the overdrive technique with a whole surface can be difficult. First, it is almost impossible to get a good transfer function for the entire mirror since the response will not be the same depending on the spatial frequency of the aberration. One would need to compute a transfer function for each Zernike polynomial and then decompose the target aberration into those polynomials. But at the end, the mirror would be limited by the highest spatial frequency that can be corrected. Instead, we chose not to compute a transfer function for the whole mirror but to try to do it experimentally. Even then, an important issue arises: since the settling time is indirectly related to the amplitude because of the limits of the equipment that we use, as explained before, the speed of the mirror will be limited by the actuator that needs the most current (the highest amplitude on the surface). In that case, it is useless to perform the same optimization procedure employed for a single actuator (finding the transfer function, simulating the best settling time and fine tuning the parameters experimentally) and do it for all 37, which would be a time-consuming process. Instead, we chose to look at the most demanding (in current) actuator and find the ratio of overdrive and target command from there. After computing the current from the influence matrix [1], the highest current was identified to fix the overdrive and the overdrive time was subsequently fine- tuned experimentally. Figure 9 shows the experimental results obtained for a defocus.

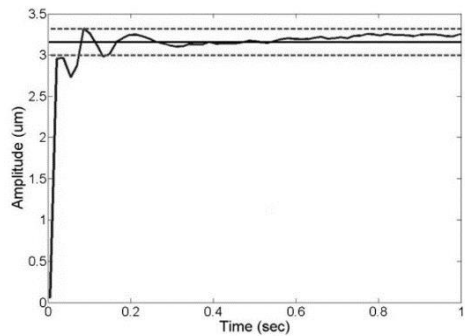

Figure 9. Experimental result of an overdrive command creating the aberration of defocus in the 37-actuators configuration. The dotted lines give the $\pm 5\%$ margin. The settling time is $100 ms$. This is a single measurement, not averaged.

Considering a $\pm 5\%$ margin [8], the settling time with the overdrive for the aberration of defocus is $100\ ms$. If this margin is increased, it could be shortened to $50\ ms$. Earlier, when considering a single actuator, the wavefront shape was not a concern since it was only the step response of a single actuator. In the case of an entire surface, the shape must be considered. Figure 10 shows the wavefront measured at $100ms$.

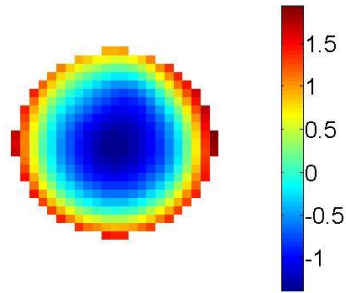

Figure 10. Experimental wavefront taken at $t = 100ms$ from Figure 9. The amplitude shown on the color bar is in microns. This is a single measurement, not averaged.

The wavefront is not perfect and for this reason, it is important to also consider the error as a function of time when dealing with a whole surface. Figure 11 shows the evolution of the peak-to-valley error as a function of time.

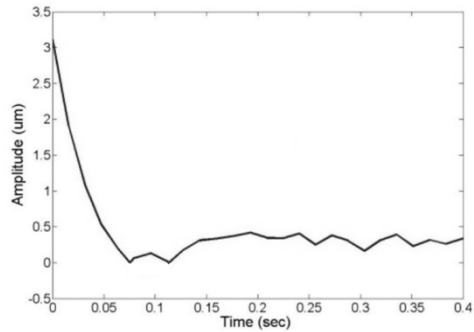

Figure 11. Experimental peak-to-valley error from Figure 9. Each wavefront is subtracted from the theoretical aberration and the evolution of the amplitude is then plotted. This is a single measurement, not averaged.

It is much harder to optimize the controls since we need to take care of both curves (Figure 10 and Figure 11) at the same time. The last technique we present here concerns the use of an elastomer membrane placed over the ferrofluid. The deposition itself is difficult for there must be no air gap between the liquid and the membrane. When this is achieved, we then need a stronger magnetic field to deform the ferrofluid since the membrane will decrease the amplitudes of the deformations. We therefore made a bigger actuator with more wire turns and wires having larger diameters so that it can handle higher currents. The purpose of this experiment was only to characterize the dynamics of the mirror with the membrane. In practice, we needed the big actuator so that we could use the overdrive controller on the mirror, because the mirror that we presently have cannot generate a high enough deformation with a low current. With our present mirror, we must drive the actuators to their maximum limit only to have a few microns of deformation while using the membrane. To use the elastomer membrane, we also needed to make a homemade water-based ferrofluid because the membrane needs a hydrophilic fluid. This homemade ferrofluid has a higher viscosity and a higher density than the EFH1 ferrofluid used in the other experiments so that its dynamics is completely different. In particular, the settling time is considerably longer as shown in Figure 12.

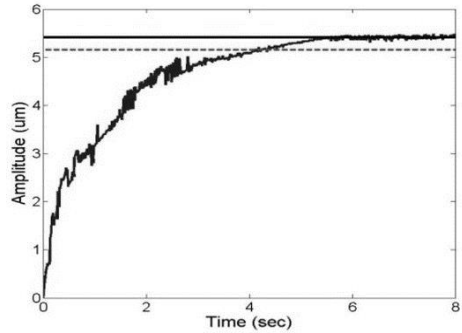

Figure 12. Experimental result of a step response with a bigger unique actuator and the homemade ferrofluid. This characterization is done as a reference before laying down the elastomer membrane. The settling time is about $4sec$. This is a single measurement, not averaged.

Figure 12 shows that the settling time of this homemade ferrofluid is about $4sec$. The target amplitude was $5.2\mu m$. Figure13 shows the experimental result with the membrane deposited on the ferrofluid.

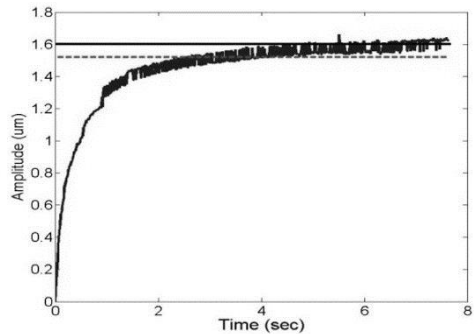

Figure 13. Experimental result of a step response with a bigger unique actuator and a homemade ferrofluid using the elastomer membrane. The settling time has come down to about $2sec$. This is a single measurement, not averaged.

By comparing Figure 13 and Figure 12, we see a few interesting results. First it is clear that using the membrane decreases the amplitude by about a factor of 3 as both figures were taken using the same target amplitude of $5.2\mu m$. The most important improvement is that there is a gain in settling time, from $4sec$ to $2sec$, even without using the overdrive controller. Then, the big actuator is used to send an overdrive command on the homemade ferrofluid without a membrane to see what kind of gain in speed is achievable. Figure 14 shows the experimental results.

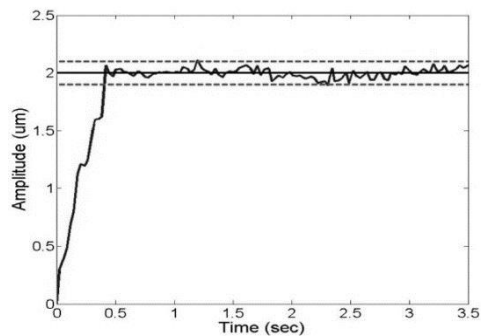

Figure 14. Experimental result of an overdrive command using the high viscosity and high density homemade ferrofluid without the elastomer membrane for a single

actuator. The dotted lines are the ±5% margin and the settling time is down to $400ms$. This is a single measurement, not averaged.

Figure 14 shows that the settling time was improve by a factor of 10, from $4sec$ to $400ms$. Doing this with the EFH1 is presently impossible because it is not compatible with the membrane. However, this result shows that the settling times obtained with the EFH1 ferrofluid in the experiments used to make the previous figures could be improved by a factor of 10 by using a comparable hydrophilic ferrofluid. Finally, we carried out experiments using both the membrane and the overdrive controller.

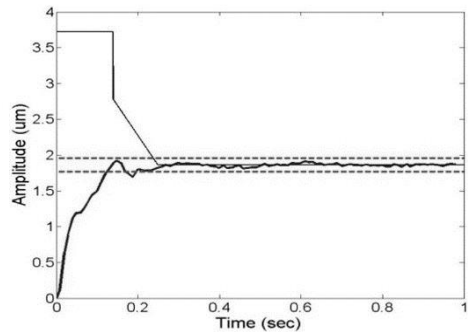

Figure 2. Experimental result of an overdrive command using the high viscosity and high density homemade ferrofluid with the elastomer membrane for a single actuator. The dotted lines give the ±5% margin, the solid square line is the evolution of the command and the settling time is down to $200ms$. This is a single measurement, not averaged.

Figure 15 shows the experimental results. Comparing them to those in Figure 13, we see that the settling time is decreased by a factor of 20, from **$4sec$** to **$200ms$**.

**Conclusion**

After having characterized the mirror spatially in previous publications (1-5), we needed to improve its time response and its reflectivity. After calculating the transfer function of the system, it was possible to make some simulations and come out with the best controller to use to reduce the settling time. The overdrive controller allowed us to decrease the settling time from $100ms$ to $25ms$ using a factor of 10 ratio between overdrive and target commands. The settling times could be decreased even more by increasing this ratio with the next generation mirrors. This can be seen by comparing Figure 6, obtained from computer simulations, to Figure 4. We also carried out the same process with a whole surface and found that the results can be optimized with a better ratio of commands.

Adding an elastomer membrane to the surface of the mirror improved the time response, the reflectivity and the influence function. Experimental results show that the membrane decreased the settling time by a factor of 2. While there is still work to be done to develop a hydrophilic ferrofluid with low viscosity, we can see an improvement of a factor of 20 in settling times using the overdrive controller and the reflective membrane. Because the computer simulations in Figure 6 show a settling time of $5ms$, we see that the next generation of ferrofluidic mirrors coated with a reflective membrane could reach settling times of the order of a millisecond. Presumably even lower settling times could be possible. As explained in [1], the use of a Maxwell coil around the mirror helps linearizing the response of the actuators but also increases the amplitude of the deformation. By actively controlling this Maxwell coil, one could also apply an overdrive controller on it. This might imply to add a cooling system around the mirror or add thermal insulation in between the actuator or the ferrofluid to make sure not to be affected by convection current.

The loss of amplitude that comes from the use of a membrane is not a critical problem since very high amplitudes have been demonstrated with ferrofluidic mirrors [1]. Finally, the membrane also narrowed the influence function by 15% which decreases the inter-actuator coupling allowing them to be closer which gives more degrees of freedom in correcting aberrations.

**References**

1. D. Brousseau, E.F. Borra, M. Rochette and D. Bouffard-Landry, *Linearization of the response of a 91-actuator magnetic liquid deformable mirror.* Opt. Express **18**(8), 8239-8250 (2010).
2. J. Parent, E.F Borra, D. Brousseau, A.M. Ritcey, J.-P. Déry and S. Thibault, *Dynamic response of ferrofluidic deformable mirrors*, Appl. Opt. **48**(1), 1–6 (2009).
3. P. Laird, E.F. Borra, R. Bergamesco, J. Gingras, L. Truong and A.M. Ritcey, *Deformable mirrors based on magnetic liquids*, Proc. SPIE **5490**, 1493–1501 (2004).
4. D. Brousseau, E.F. Borra, and S. Thibault, *Wavefront correction with a 37-actuator ferrofluid deformable mirror*, Opt. Express **15**(26), 18190–18199 (2007).
5. D. Brousseau, E.F. Borra, S. Thibault, A.M. Ritcey, J. Parent, O. Seddiki, J.-P. Dery, L. Faucher, J. Vassallo and A. Naderian, *Wavefront correction with a ferrofluid deformable mirror: experimental results and recent developments*, Proc. SPIE **7015**, 70153J (2008).
6. D.Brousseau, J.-P. Véran, S. Thibault, E.F. Borra, S. F.-Boivin, *Woofer-tweeter adaptive optics in very strong turbulence using a magnetic-liquid deformable mirror*, Proc. SPIE **8447**, 84473Z (2012).
7. R.J. Noll, Zernike polynomials and atmospheric turbulence, J. Opt. Soc. Am, **66**(3), (1976)
8. K. Ogata, *Modern Control Engineering*, **5**$^{th}$ edition, Pearson, (2010).
9. A. Iqbal, and F. B. Amara, Modeling of a Magnetic-Fluid Deformable Mirror for Retinal Imaging Adaptive Optics Systems, Int. J. Optomechatronics **1**(2), 180–208 (2007).